\documentclass[fleqn,useAMS,usenatbib]{mn2e}
\usepackage[suffix=]{epstopdf}
\usepackage{graphicx}
\usepackage{natbib}
\usepackage{amsmath}
\usepackage{url}
\usepackage{times}

\def\lesssim{\mathrel{\hbox{\rlap{\hbox{\lower3pt\hbox{$\sim$}}}\hbox{\raise2pt\hbox{$<$}}}}}
\def\gtrsim{\mathrel{\hbox{\rlap{\hbox{\lower3pt\hbox{$\sim$}}}\hbox{\raise2pt\hbox{$>$}}}}}

\usepackage{aas_macros}

\providecommand{\e}[1]{\ensuremath{\times 10^{#1}}}

\title[Ten Dimensional Stellar Locus]{The SDSS--2MASS--WISE Ten Dimensional Stellar Color Locus}

\author[James R. A. Davenport et al.]{James R. A. Davenport,$^{1}$\thanks{E-mail: jrad@astro.washington.edu} \v{Z}eljko Ivezi\'{c},$^{1}$ Andrew C. Becker,$^{1}$ John J. Ruan,$^{1}$
\newauthor  Nicholas M. Hunt-Walker,$^{1}$ Kevin R. Covey,$^{2}$ Alexia R. Lewis,$^{1}$
\newauthor  Yusra AlSayyad,$^{1}$ Lauren M. Anderson$^{1}$\\
$^{1}$Department of Astronomy, University of Washington, Box 351580, Seattle, WA 98195\\
$^{2}$Lowell Observatory, Flagstaff, AZ 86001, USA
}

\begin{document}

\pagerange{\pageref{firstpage}--\pageref{lastpage}} \pubyear{2013}
\maketitle
\label{firstpage}

\begin{abstract}
We present the fiducial main sequence stellar locus traced by 10 photometric colors observed by SDSS, 2MASS, and WISE. Median colors are determined using 1,052,793 stars with $r$-band extinction less than 0.125. 
We use this locus to measure the dust extinction curve 
relative to the $r$-band, which is consistent with previous measurements in the SDSS and 2MASS bands. The WISE band extinction coefficients are larger than predicted by standard extinction models. 
Using 13 lines of sight, we find variations in the extinction curve in $H$, $K_s$, and WISE bandpasses. Relative extinction decreases towards Galactic anti-center, in agreement with prior studies. Relative extinction increases with Galactic latitude, in contrast to previous observations. This indicates a universal mid-IR extinction law does not exist due to variations in dust grain size and chemistry with Galactocentric position. A preliminary search for outliers due to warm circumstellar dust is also presented, using stars 
with high signal-to-noise in the W3-band. 
We find 199 such outliers, identified by excess emission in $K_s-W3$. Inspection of SDSS images for these outliers reveals a large number of contaminants due to nearby galaxies. Six sources appear to be genuine dust candidates, yielding a fraction of systems with infrared excess of $0.12\pm0.05$\%.
\end{abstract}

\section{Introduction}
The Two Micron All Sky Survey \citep[2MASS;][]{2mass} and Sloan Digital Sky Survey \citep[SDSS;][]{york2000} have provided revolutionary improvements in our understanding of the stellar populations within our Galaxy at near infrared and optical wavelengths, respectively. For example, Milky Way halo substructures have been discovered in both SDSS and 2MASS photometry \citep{ibata2001,ibata2002}. Normal stars have been separated from more exotic objects with much greater accuracy by matching sources between these surveys \citep{finlator2000}. This wide-field dataset has continued to set the standard for multi-wavelength studies, enabling science not possible with either survey individually, and providing astrometric and flux standards to calibrate future surveys.

\citet[][hereafter C07]{covey2007} used a sample of $\sim$600,000 stars matched between SDSS and 2MASS to measure the fiducial stellar color locus in $ugrizJHK_s$ passbands as a function of $g-i$ color. This parameterization provides colors of main sequence stars as a function of their effective temperature. 
The C07 main sequence locus has been used to search for color outliers due to being, for example, white dwarf binaries, quasars, or post-main sequence stars. The C07 locus has also provided a robust method to identify and classify normal stars given any combination of SDSS and 2MASS colors.

The Wide-field Infrared Survey Explorer \citep[WISE;][]{wise} has produced a modern census of the entire sky with unprecedented accuracy and depth in four bandpasses, ranging from 3.4 to 22 $\mu$m. Combining this survey with the well studied SDSS and 2MASS datasets will enable the discovery of new classes of both Galactic and extra-Galactic objects. Furthermore, this will provide improved understanding of the cool solar neighborhood, young stellar populations, dust content, and substructure within our Galaxy.  Already, WISE and SDSS data have been used to survey infrared excesses around white dwarfs \citep{debes2011} and to discover many new brown dwarf candidates \citep{aberasturi2011}. The first confirmed Y0 dwarf \citep{cushing2011} has also been discovered with WISE, probing the stellar mass function at its lowest extrema for the first time. WISE photometry will also provide the best dataset for mapping asymptotic giant branch, with the ability to trace them to well beyond the Galactic center (Hunt-Walker 2014 in prep). All of these studies will critically depend on the proper understanding of photometric behavior for ``normal'' stars simultaneously in all 10 colors, as was done with the seven colors from SDSS and 2MASS in C07. 

In this paper we present the first detailed study of the stellar locus for nearby stars as observed by 2MASS, SDSS, and WISE. In \S\ref{sec:data} we describe the creation of a matched sample of low-extinction point sources. A detailed measurement of the stellar locus is given in \S\ref{sec:locus}. Using this fiducial stellar color sequence as a set of ``standard crayons'' \citep[termed by][]{peek2010}, we measure the relative dust extinction coefficients from the $u$-band to 22$\mu$m in \S\ref{sec:extinct}. We search for warm dust disks from WISE color outliers in \S\ref{sec:dust}. A summary of our work is given in \S\ref{sec:conclusion}.

\section{Data}
\label{sec:data}

Our data come from three surveys that differ greatly in their sky coverage, photometric depth, and wavelength coverage. In the following section we briefly describe each of the three photometric surveys, as well as the quality cuts, selection criteria, and matching used to produce a clean sample of nearby stars to use in measuring the fiducial stellar locus.

\begin{figure}
\centering
\includegraphics[width=3.5in]{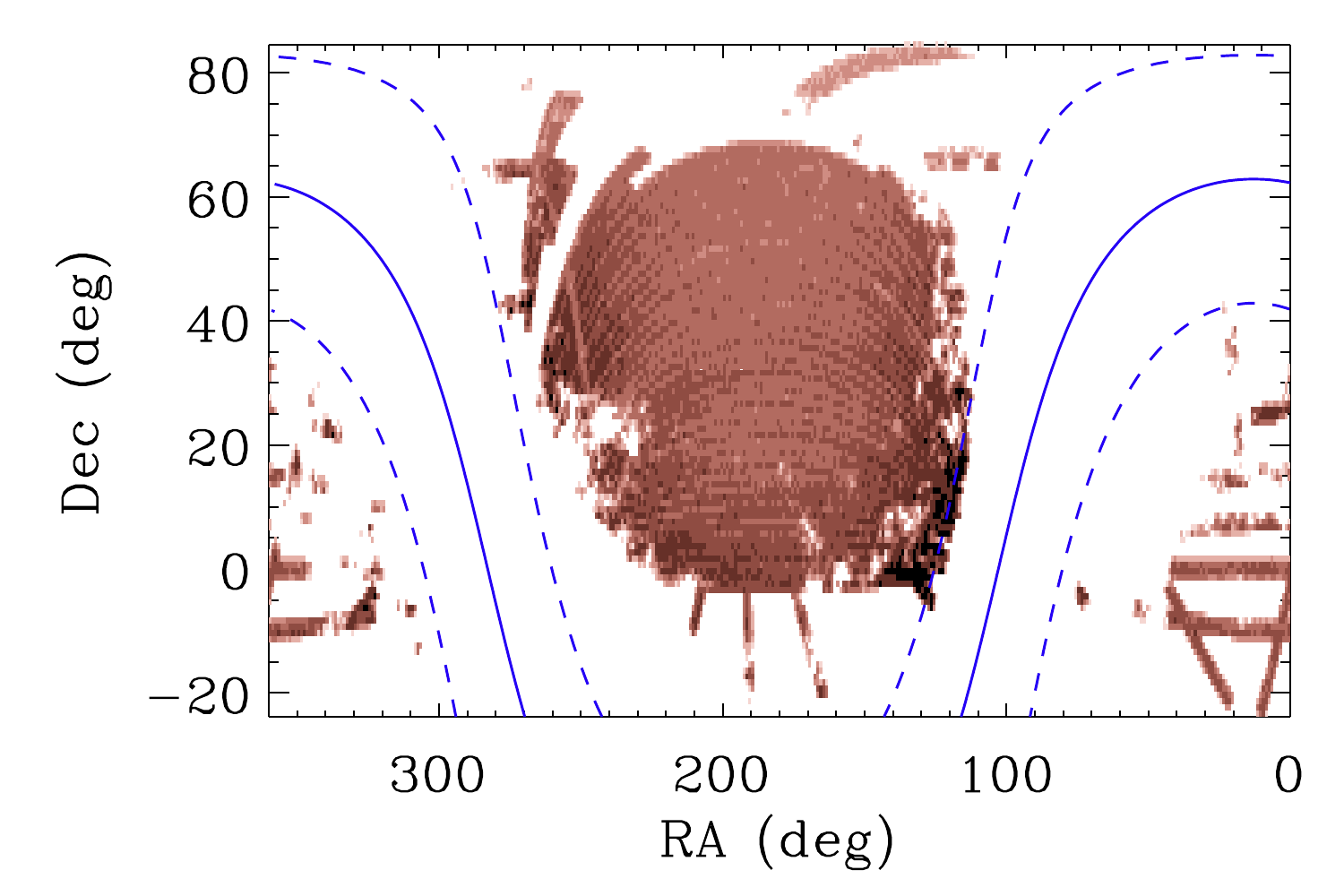}
\caption{Map of the final 1,052,793 source sample matched between the three surveys. A pixel size of 1 deg$^2$ was used, with number density of sources increasing from light to dark. 
Also shown is the $b=0$ Galactic equator (blue solid line) and $b=\pm10^\circ$ (blue dashed lines).}
\label{radec}
\end{figure}

\subsection{SDSS Photometry}
The optical photometry for our study came from the SDSS Data Release 8 \citep[DR8;][]{dr8a}. This survey contiguously mapped over 14,000 deg$^2$ of the northern and southern Galactic caps, with small extensions of the footprint through the Galactic plane. The drift-scanning technique used by the SDSS camera produced nearly simultaneous 54 second exposures in five optical filters ($u,g,r,i,z$), spanning wavelengths from 0.36 $\mu$m to 0.90 $\mu$m \citep{fukugita1996}.  The DR8 photometry has a nominal calibration precision of 1\% for $griz$-bands, and 2\% for the $u$-band.

We used the value-added tables of SDSS DR8 data provided from \citet{berry2012}, which included an independent extinction measurement for each source using SDSS and 2MASS photometry. 
We selected only photometric objects with the object flag set to Objc\_Type=6, which corresponds to objects the SDSS pipeline determined to be unresolved point sources (dominated by stars and quasars). The error rate of this morphological classification is below 5\% at $r=21$ \citep{lupton2001}. We used standard SDSS flag cuts to select high quality point sources, including BINNED1=1, BRIGHT=0, SATURATED=0, EDGE=0, BLENDED=0, and NOPROFILE=0.

We required each object to have $13.8<r<21.5$. To remove the majority of quasars and white dwarfs from our sample we selected objects with $u-g > 0.6$. Following C07, to select main sequence stars we choose objects within the range $-0.1<g-i<5$.

Since interstellar dust causes wavelength-dependent extinction and reddening, we selected objects with very low extinction $A_r\le0.125$, as provided by both the DR8 database and \citet{berry2012}.
The DR8 database extinction estimates are derived from the dust maps by \citet{sfd}, and were verified as accurate for low extinction regions in the SDSS by \citet{schlafly2011} and \citet{berry2012}. This criterion was more strict than the $A_r\le0.2$ used by C07, and ensured a sample of stars where effects of extinction errors would be much smaller than the intrinsic scatter in the locus.

\subsection{2MASS and WISE Photometry}
The mid-infrared photometry we used came from the latest AllWISE data release. Simultaneous photometry was provided in four filters, with central wavelengths of 3.4, 4.6, 12, and 22 $\mu$m. Astrometric calibration for WISE was achieved by matching sources to the full-sky 2MASS point source catalog \citep{wise}. As a result, the WISE  database provides 2MASS $JHK_S$ photometry for every source that had a match between these surveys. An independent cross match between these two surveys was therefore not necessary.

We required every source to have a detection in the $J$-band and W1-band, and selected point sources with $W1<13.6$ mag to remove objects fainter than the WISE completeness limit. Photometric uncertainty requirements of  $\sigma_{W1}<0.05$ mag, $\sigma_{W2}<0.1$ mag, and $\sigma_{W3}<0.2$ mag were also imposed to ensure 5-$\sigma$ detections in WISE bands. Spurious detections of sources were removed by requiring each band to have a photometric error $>0$. A rough method to remove nearby giant stars from the sample was used by selecting sources with $J>12$ \citep{covey2008}.

\subsection{Matched Sample}
The AllWISE data covered the entire SDSS DR8 footprint, including the low Galactic latitude stripes. We positionally matched these two databases using a radius of 2\farcs0. This retrieved 9.9\e{6}unique objects with a detection in SDSS and at least a 5-$\sigma$ detection in the W1 and W2 bands.

Imposing the SDSS quality cuts, as well as selecting sources with $A_r<0.125$, yielded 5.4\e{6} matched sources. The final sample used to measure the stellar locus, after all quality and selection cuts in both optical and infrared bands, and removing sources with low Galactic latitude $|b|<10$, contained 1,052,793 stars. The sky distribution of this matched sample is shown in Figure \ref{radec}. For measurements in the W3 we placed additional constraint on the sample, requiring $W3<11.5$ and $W3$ PH\_QUAL = A or B This produced a W3 sample of only 7,430 stars. For the $W4$ band no stars were able to pass the requirement of $W4$ PH\_QUAL = A or B. As a result we cannot trace the detailed stellar locus in $W4$ with this sample.

The SDSS DR8 data were calibrated to the AB photometric system \citep{abmag}, while the WISE and 2MASS data have been calibrated to the Vega system \citep{cohen1992,cohen2003}. As in C07, we have used the data from each survey in its native magnitude system, mixing AB and Vega magnitudes.

\begin{figure}
\centering
\includegraphics[width=3.5in]{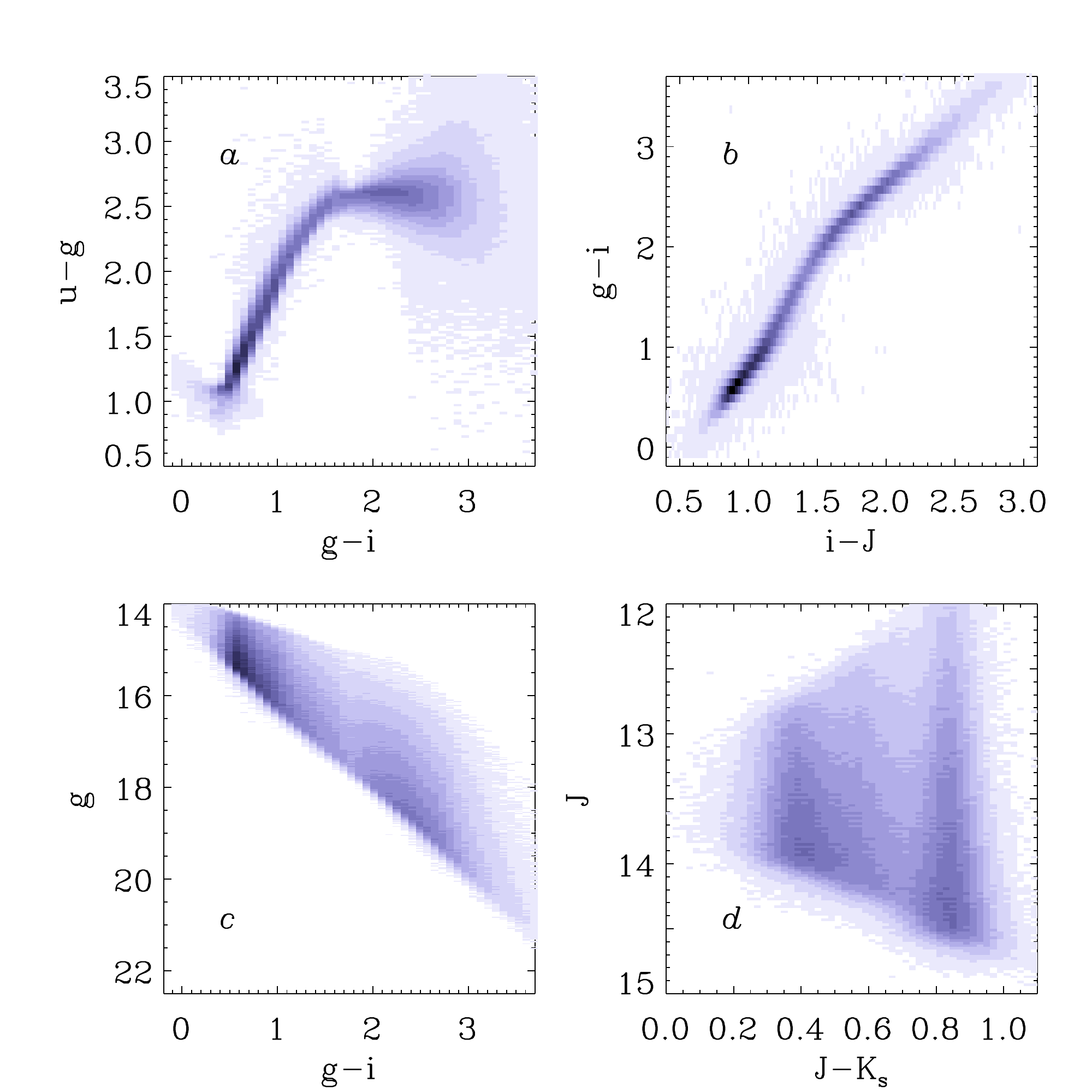}
\caption{Color vs. apparent magnitude and color vs. color plots for the SDSS passbands of our low-extinction matched sample. Pixel intensity from dark to light represents an exponentially increasing density of stars.}
\label{sdss}
\end{figure}

In Figure \ref{sdss} we show a subset of the color--magnitude and color--color diagrams for our matched sample in the SDSS and 2MASS bands. The $(g-i, u-g)$ diagram can be used to identify many types of stars, binaries, and white dwarfs, as well as contaminants from quasars \citep[e.g.][]{ivezic2007}. The position along the narrow locus in the $(i-J,g-i)$ diagram of Figure \ref{sdss} 
is parameterized by effective temperature over a wide range of stellar mass, as shown by 
\citet{bochanski_templates} and \citet{westdr7} for red stars, and \citet{ivezic2008} for blue stars.
The SDSS $(g-i,g)$ color--magnitude diagram in Figure \ref{sdss} reveals that our sample is almost entirely composed of stars from the Galactic thin disk at distances $\lesssim 1$ kpc \citep{ivezic2008}.

\section{The 10-Dimensional Locus}
\label{sec:locus}

Using the 11 passband sample defined above we measured the median colors of stars in all 10 adjacent photometric colors. As this sample was selected on the basis of low extinction, we did not apply reddening corrections. C07 showed that for main sequence stars the $g-i$ color traced effective temperatures 3540 K $< T_{eff} <$ 7200 K. 
We measured the color locus in each of the filter combinations $(u-g,g-r,r-i,i-z,z-J,J-H,H-K_s,K_s-W1,W1-W2,W2-W3)$ as a function of their $g-i$ color. 

\begin{figure*}
\centering
\includegraphics[width=6.5in]{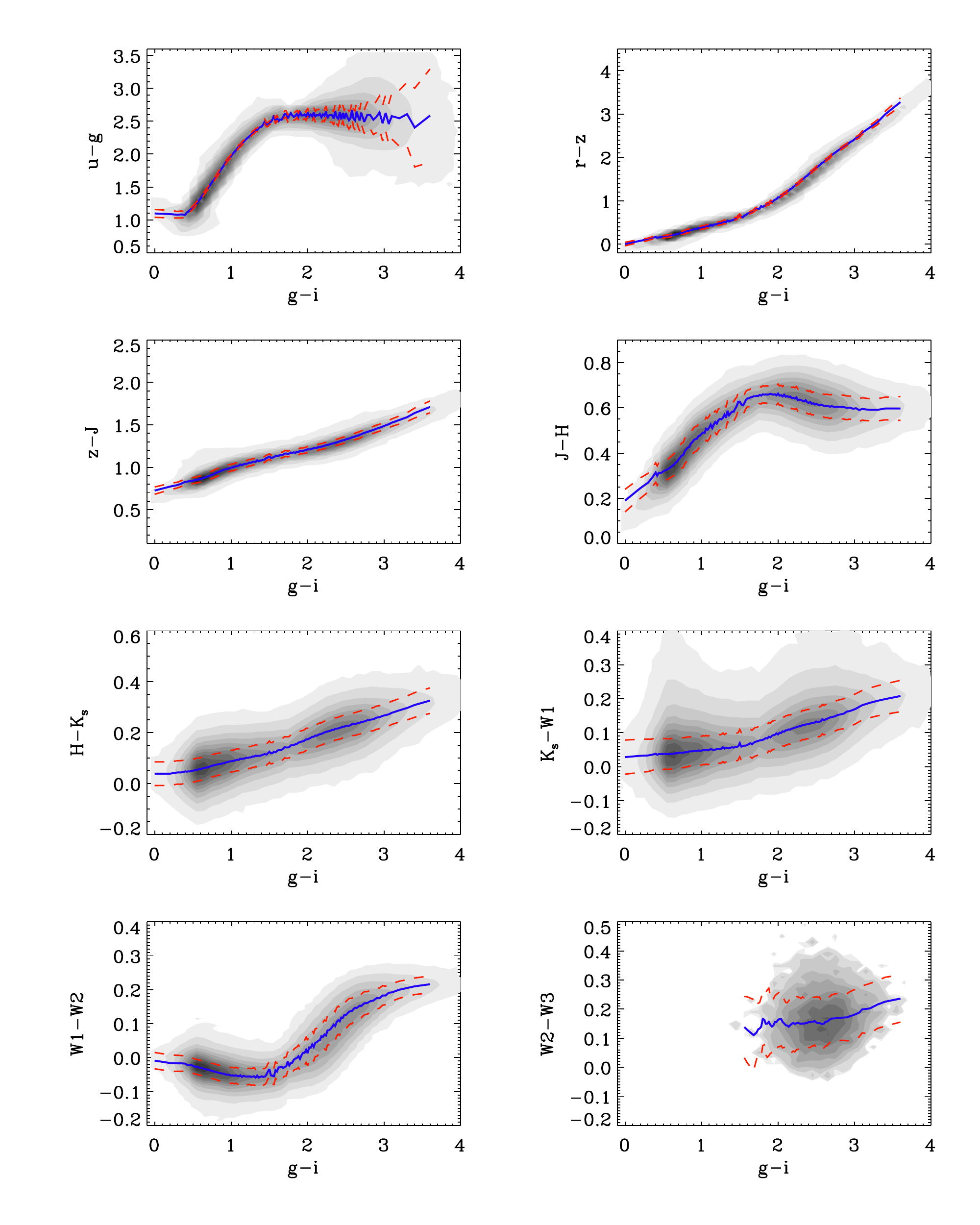}
\caption{Subsets of the median 10-D color locus (blue line) as a function of $g-i$ color. The standard deviation measured in each $g-i$ bin is also indicated (red dashed line). Shaded density pixels show our low-extinction matched sample, with source density increasing exponentially from light to dark.}
\label{locus}
\end{figure*}

Figure \ref{locus} shows a subset of our measured median color locus. 
Following C07, the locus was defined in small $\delta(g-i)$ steps. Because our sample selection placed strong limitations on both extremely blue and red objects, the bin size was increased near both ends of the locus to increase the numbers of stars per bin.
A bin size of $\delta(g-i)=0.02$ mag was used from $0.4<g-i<2.7$, and was increased to $\delta(g-i)=0.04$ mag from $2.7<g-i<3.1$, and then further increased at both the red and blue limits.  The median and standard deviation of all 10 colors was computed within each $\delta(g-i)$ bin, tracing out the stellar locus (and its width) in the adjacent colors. These measurements, along with the number of stars for each $\delta(g-i)$ bin, are given in Table \ref{bigtable1} for SDSS bands, and Table \ref{bigtable2} for 2MASS and WISE bands. 
There was an average of 2000 stars per color bin, which is sufficient for statistical errors to be negligible compared to the systematic photometric uncertainties. 
However, we did not have a sufficient number of blue stars that passed our $W3$-specific selection criteria. As a result we only provide the $W2-W3$ locus for red stars with $g-i>1.4$. To smooth out non-physical variations in the locus due to small sample sizes we applied a boxcar smoothing kernel to the $W2-W3$ locus, using a kernel width of $g-i$=0.1. 
Excellent agreement with the C07 measurement is found for the SDSS and 2MASS colors, with a median difference between the two loci of 0.015 mag.

\begin{figure*}
\centering
\includegraphics[width=6in]{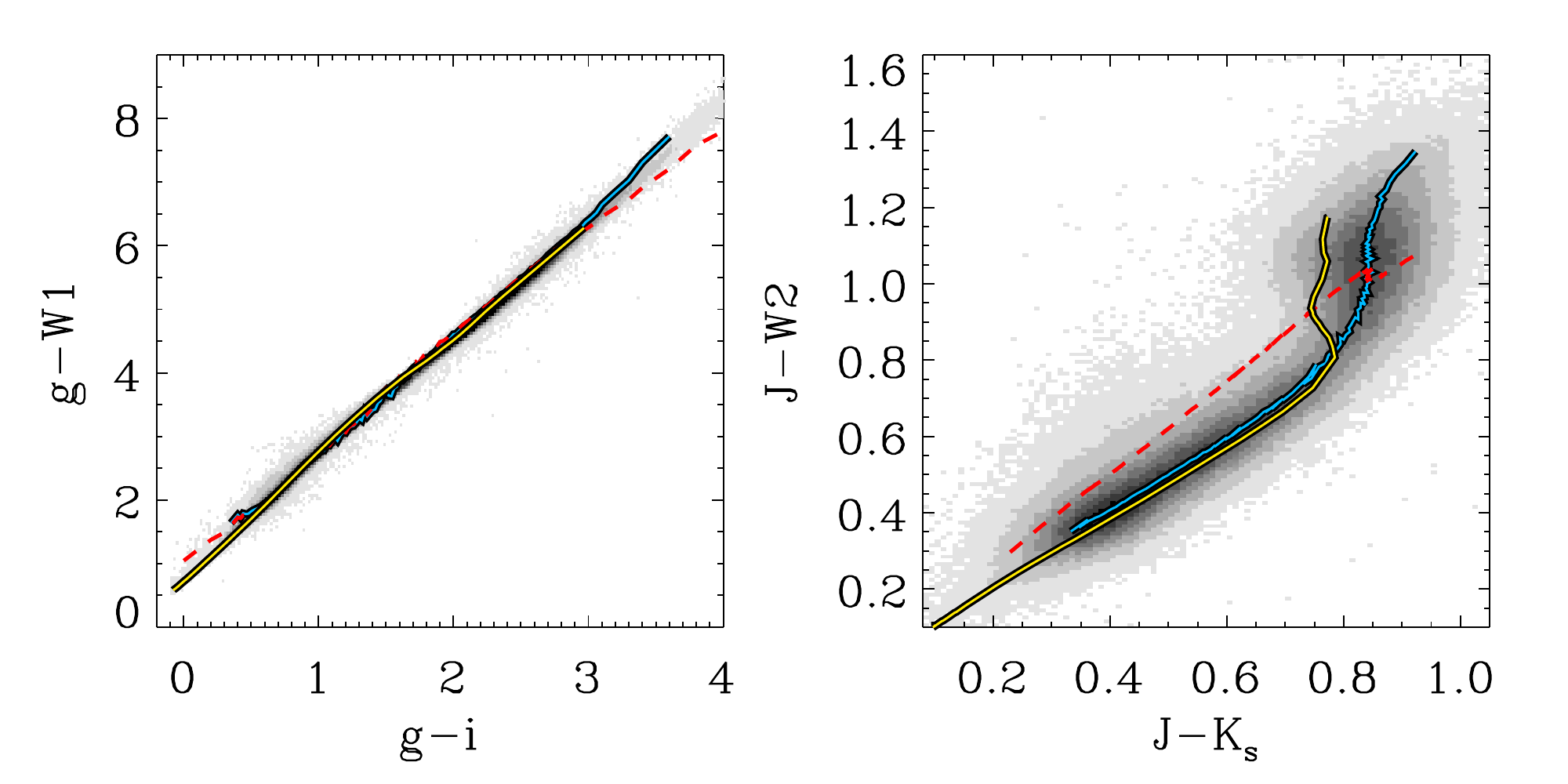}
\caption{Comparison of our locus (blue line) to the \citet{bilir2011} fits (red dashed line), using the colors provided in \citet{bilir2011}. We also show the \citet{bressan2012} isochrone model (yellow line) using Z=0.019 and t=10$^{9}$ years. Shaded pixels in the background show our low-extinction matched sample, with density increasing from light to dark.}
\label{bilircomp}
\end{figure*}

\subsection{Locus Width}
The width of the stellar locus about the median track in Figure \ref{locus} varies notably between panels. This is due to the different photometric errors in each bandpass. The increasing width of the $u-g$ color for low-mass stars ($g-i>2$) for example is due to errors in the $u$-band measurement for these red objects. For the 2MASS and WISE colors, however, the locus width is consistent with scatter due to photometric errors. Dividing the locus standard deviation by the median photometric color error in each $(g-i)$ bin, we find the weighted scatter in the locus ranges from values of $\sim$0.9 to $\sim$1.2, consistent with Poisson noise. Our results for the locus width also compare very favorably to C07. The median difference in the locus standard deviations for the SDSS and 2MASS colors between C07 and our study was 0.011 mag.

\subsection{Comparison to Previous Loci}
As mentioned above, we find a median difference of 0.015 mag between our locus values and that of C07, smaller than the $g-i$ bin size used in either study. In Figure \ref{bilircomp} we compare our stellar locus to a previous determination of the colors of 825 nearby stars by \citet{bilir2011}. We used their Equations 3 and 6, with the $(g-r), (r-i), (J-H),$ and $(H-K_s)$ values from our locus, to generate direct comparisons in the color spaces they provide. We also show the 1 Gyr  isochrone model, with metallicity of Z=0.019, from \citet{bressan2012}. This isochrone model provides colors for main sequence stars with masses in the range of $0.1\le M_\odot \le2.15$.
Our locus differs from both the \citet{bilir2011} measurement and \citet{bressan2012} isochrone track by less than 0.1 mag in $g-W1$ color over the range $0.25<g-i<3$. However, neither the \citet{bilir2011} or \citet{bressan2012} tracks accurately reproduce the $J-W2$ color space shown in Figure \ref{bilircomp}.

\section{Extinction Law}
\label{sec:extinct}

Infrared extinction coefficients (e.g. $A_{W3}$) are expected to be smaller than those of optical bands (e.g. $A_g$) by an order of magnitude \citep{schlafly2011}. \citet{bilir2011} generated extinction estimates for WISE passbands by interpolating \citet{cardelli1989} model values to the central wavelengths of the WISE filters. Recently \citet{berry2012} examined the extinction coefficients in SDSS and 2MASS filters by measuring the distortion of the stellar locus using photometry for 73 million stars in the SDSS footprint. They simultaneously determined optical and NIR extinction coefficients in each filter relative to the $r$-band, showing a significant improvement on previous extinction estimates for stars at low Galactic latitude. 
By utilizing these reliable optical extinction values, and armed with our 10-dimensional color locus, we have empirically determined the extinction coefficients in the WISE filters due to interstellar dust.

\subsection{Optical--Infrared Extinction Law}
Using the full sky AllWISE data release matched to the \citet{berry2012} SDSS DR8 with a radius $\le$ 2\farcs0, we generated a high quality matched sample to measure the 11-band relative dust extinction.  We imposed the following selection cuts:
\begin{itemize}
\item $r< 20$
\item $W1-W2 < 0.8$
\item $W1<15.8$, $W2< 14.8$, $W3 < 10.5$
\item $W2-W3 \le 2.0$
\item w1snr \& w2snr $> 5$
\item w1sigmpro \& w2sigmpro \& w3sigmpro $>0$
\end{itemize}
which produced a sample of 7.8\e{6} stars. From this sample 502,378 objects had robust $A_r$ measurements from \citet{berry2012}, with extinctions ranging from $0\le A_r\le3$. These stars were  spread across the SDSS footprint, including at low galactic latitudes. For every star in this subsample, we used the \citet{berry2012} de-reddened $g-i$ color to find its intrinsic position along our 10-D color locus.

\begin{figure}
\centering
\includegraphics[width=3in]{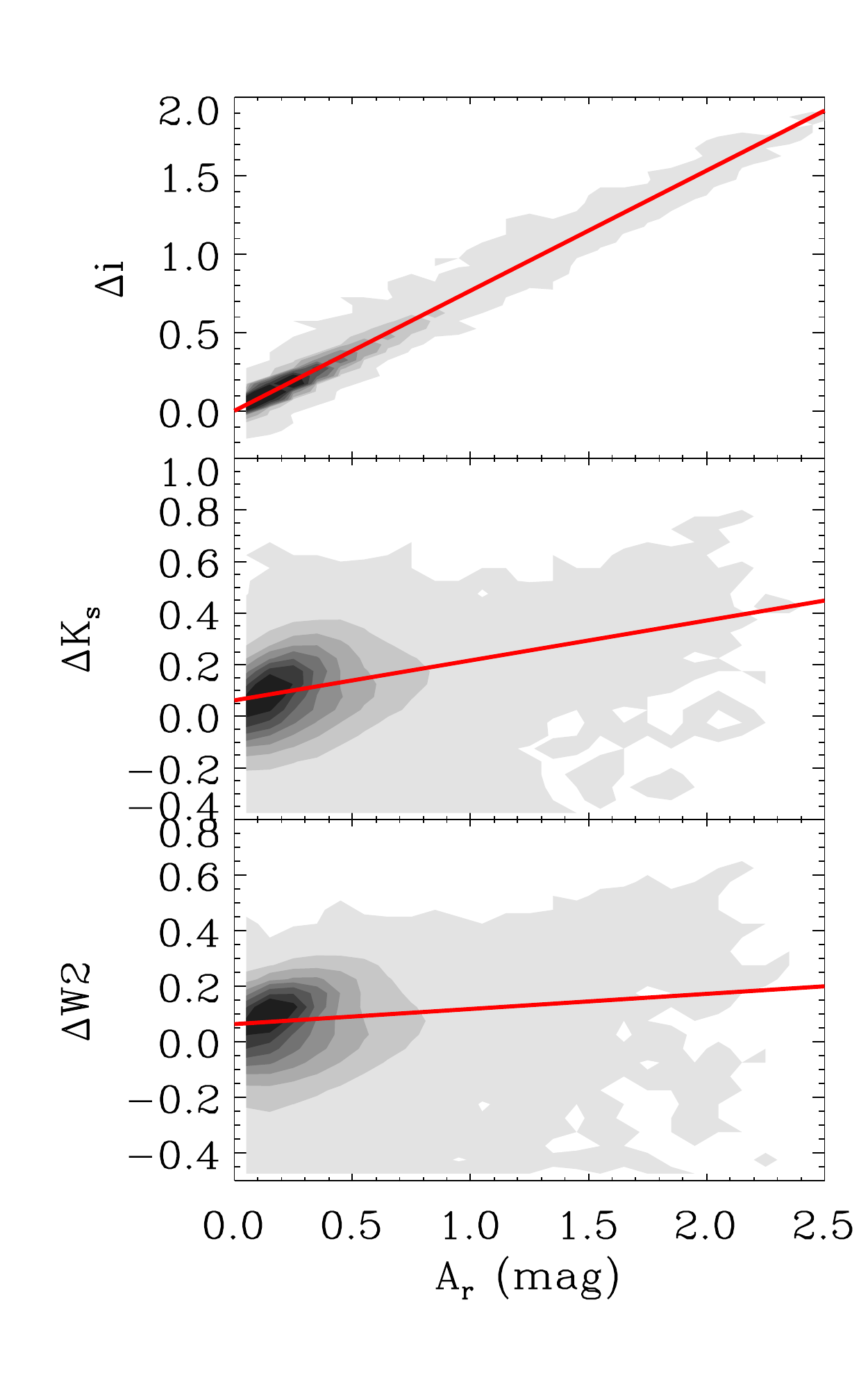}
\caption{Three examples of our method for measuring the extinction. Background contour shade indicates density of sources. The slope of the resulting linear fit (red) is $A_\lambda/A_r$.}
\label{slope}
\end{figure}

Given each star's $g-i$ position along the 10-D locus, and adopting the $r$-band extinction, we then solved for the extinction in each band relative to $A_r$. We independently measured the relative extinction values for the remaining SDSS and 2MASS bands, rather than simply adopting the \citet{berry2012} values. The basic extinction equation, using the $z$-band as an example, is:
\begin{equation}
(r-z)_{obs} = (r-z)^* + (A_r - A_z) \; ,
\end{equation}
where $(r-z)_{obs}$ is the observed color of the star, $(r-z)^*$ the intrinsic color for the star from our 10-D locus, $A_r$ is the given extinction from the SDSS database, and $A_z$ the extinction we wish to solve for. It is trivial to rearrange this equation to solve for $A_z$, which is then the only unknown term. By definition, $A_z \equiv z_{obs} - z^*$, which is equivalent to reducing Eqn. 1 given the assumption that $A_r$ and our 10-D locus is correct. Thus we computed 
\begin{equation}
A_\lambda(A_r) = \lambda_{obs} - \lambda^* = A_r - (r-\lambda)_{obs} - (r-\lambda)^*
\end{equation}
for the filters $\lambda=\{u,g,i,z,J,H,K_s,W1,W2,W3\}$. Due to the small number of systems with good $W4$ photometry in this matched sample, we were not able to provide robust measurements of the extinction in the $W4$ band. Three example panels of $A_\lambda(A_r)$ are shown in Figure \ref{slope}. This method is similar to other ``color-excess'' techniques used in determining extinctions for stars \citep[e.g.][]{gao2009,berry2012}.

The $A_\lambda(A_r)$ distributions were highly linear, with low numbers of stars showing scatter larger than the photometric errors. The slope of this linear distribution was the relative extinction coefficient in each band. This scatter accounted for less than 5\% of the sources in the subsample used, and is due to errors in the SDSS $A_r$ values used, which propagate in to using the wrong location in the 10-D locus, and possible unresolved binaries. We also implicitly assumed that the $gri$-band extinction must obey the $R_V\sim3.1$ extinction law used in \citet{berry2012} when we assigned the star a position in our 10-D locus, which may have introduced small amounts of additional scatter.

\begin{figure}
\centering
\includegraphics[width=3.5in]{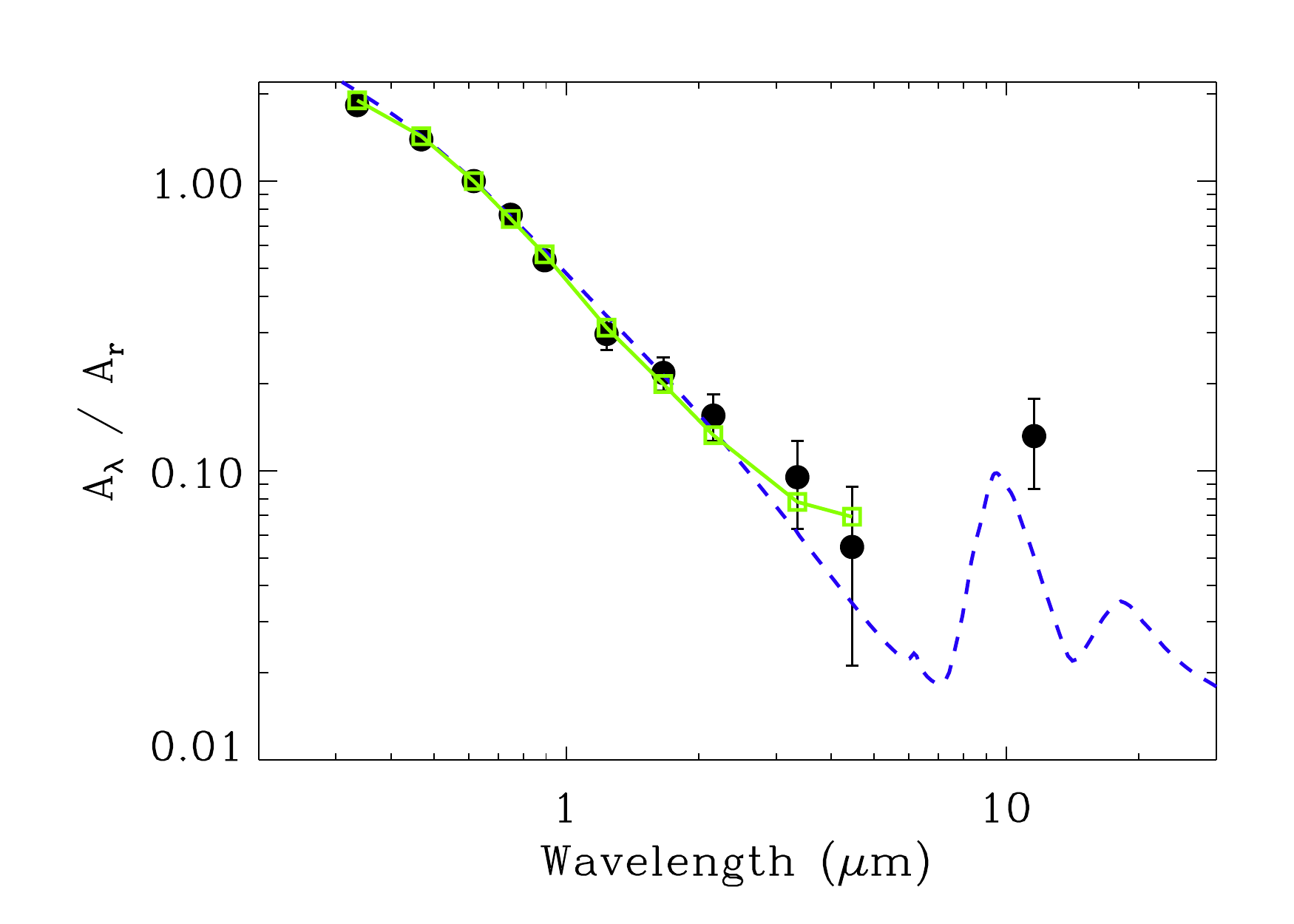}
\caption{Measured extinction coefficients $A_\lambda$ normalized to the $r$-band extinction (black circles) for our full 502,378 star sample. The \citet{wd01} $R_V=3.1$ dust model (blue dashed line) and \citet{yuan2013} empirical extinction measurements (green boxes) are shown for reference.}
\label{ck}
\end{figure}

A linear regression for each of the $A_\lambda$ versus $A_r$ distributions was computed, weighting each data point by the photometric error, using a Bayesian approach of sampling from the posterior distribution with the IDL routine LINMIX\_ERR \citep{kelly2007}. We used 5,000 iterations with the Gibbs Markov Chain Monte Carlo (MCMC) sampler for each fit. 
The extinction coefficient was determined by taking the median value of the slope from the posterior distribution of the MCMC chain. The uncertainties we list were computed as the standard deviation of the extinction laws from the 14 individual lines of sight described in the next subsection. This produced two to five times larger uncertainties than for any given extinction measurement in the infrared, and was chosen to incorporate the variations seen between different regions. These values are provided in Table \ref{extinct_table} for each filter, and are shown in Figure \ref{ck}. The empirically measured extinction coefficients from \citet{berry2012} match our measurements to within the uncertainties in every SDSS and 2MASS filter. The linear fits were not forced to go through the origin $(A_r,A_\lambda)=(0,0)$, as would be physically motivated. The best estimates from our MCMC chains were offset from the origin by less than 0.08 mag for all but the W3-band. This reddest band had the largest amount of scatter due to larger photometric uncertainties, as well as potentially blended point sources in the large W3-band aperture.

\subsection{Variations in Extinction Law}
While extinction in blue optical and UV wavelengths has the largest amplitude changes from variations in $R_V$ and $A_V$ \citep{draine2003}, the mid-IR is also a fruitful regime in which to probe the underlying properties of interstellar dust (e.g. temperature, density, grain size distribution). 
Variations in UV and IR extinction properties have been found to be uncorrelated \citep[][]{fitzpatrick2007}, indicating the lack of a ``universal'' extinction law in the mid-IR.
\citet{flaherty2007} and \citet{fitzpatrick2007} have shown that mid-IR extinction can vary significantly between different lines of sight. Further, a dependence on Galactic latitude and longitude in the shape of the IR and mid-IR extinction law, due to the structure of the Milky Way's disk and spiral arms, has also been observed \citep{fitzpatrick2009, gao2009, chen2012}.

To highlight the utility of WISE photometry in addressing such studies, we have reproduced our extinction measurements for several individual lines of sight drawn from the 502k star matched sample. We selected 10 regions corresponding to individual low Galactic latitude stripes from \citet{berry2012}. These stripes were each vertical in Galactic latitude, spaced approximately evenly in Galactic longitude from $l\sim50$ to $l\sim230$, and constrained to $|b|<25$. Using our entire 502k star sample, we also selected three non-overlapping regions in Galactic latitude, ranging from the Galactic plane with several magnitudes of extinction to high latitude regions with very little dust. These regions had limits of $|b|<25$, $25<|b|<50$, and $50<|b|<90$, all with no limits on Galactic longitude. Due to smaller sample sizes, we were not able to reliably measure extinctions in the $W3$ band in each line of sight.

\begin{figure}
\centering
\includegraphics[width=3.5in]{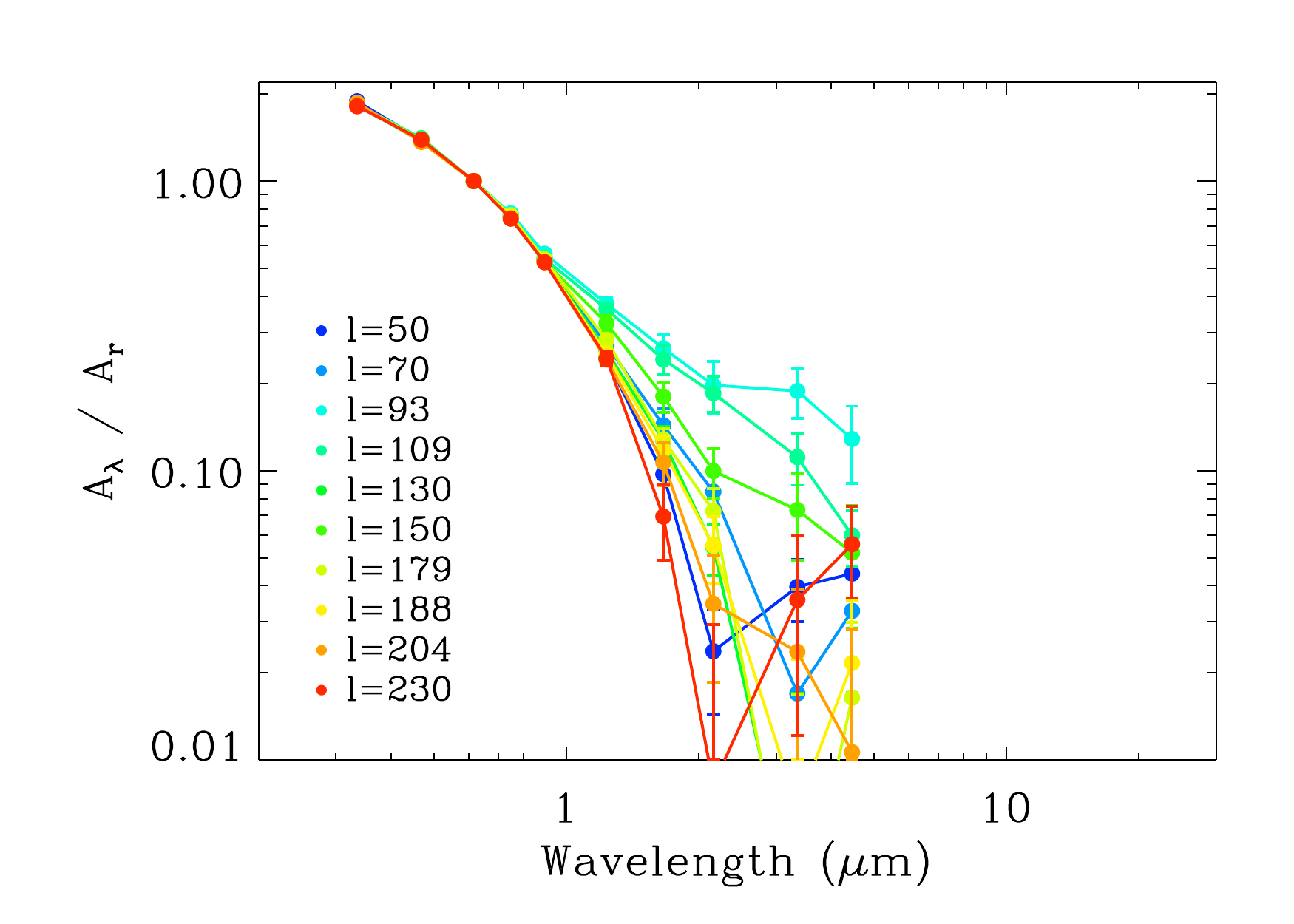}
\includegraphics[width=3.5in]{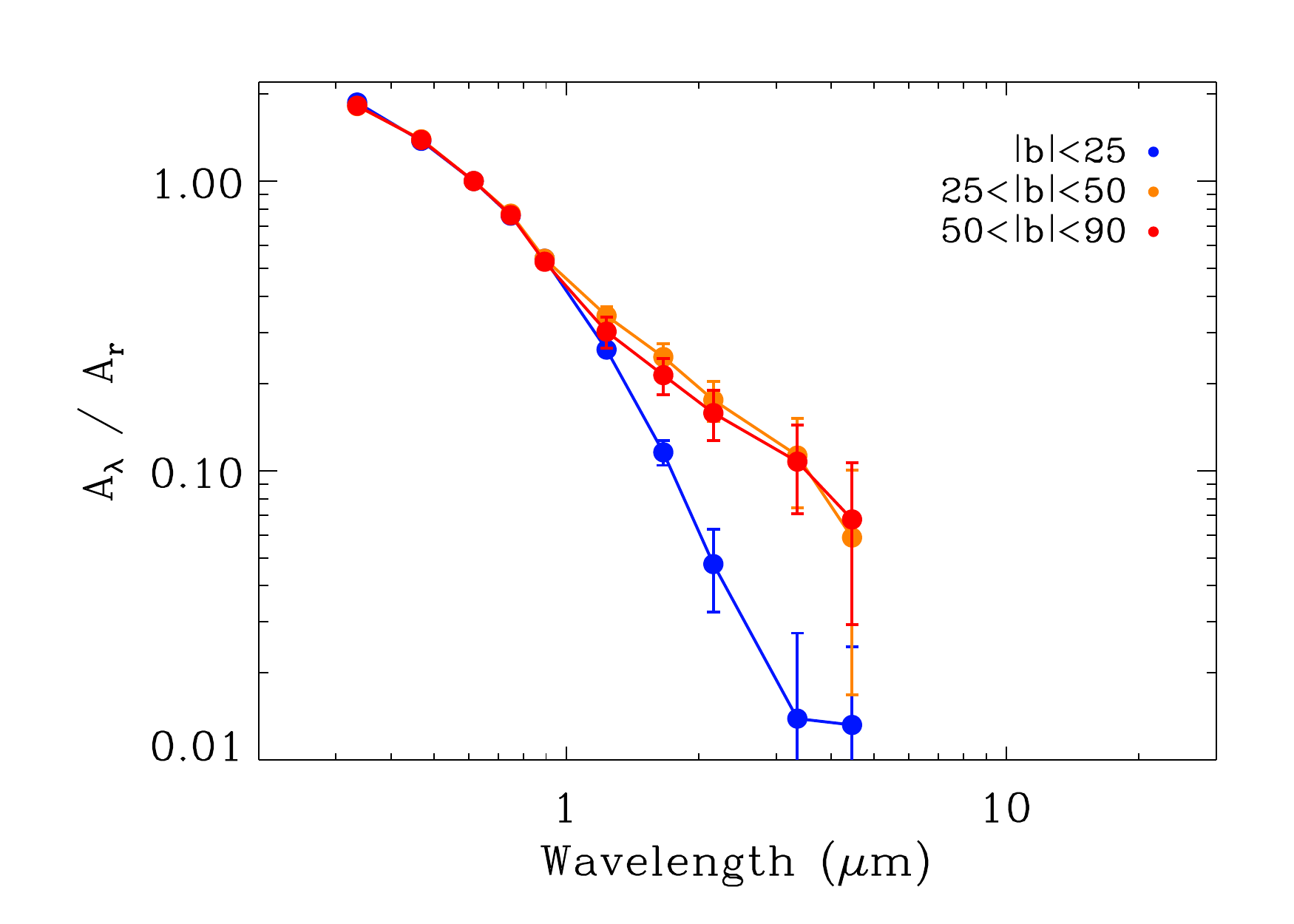}
\caption{Top: Relative extinction coefficients as a function of wavelength for ten low Galactic latitude lines of sight. Each line of sight was $\sim$3$^\circ$ wide in $l$, and limited by $|b|<25$. The median Galactic longitude for each line of sight is indicated in the legend Bottom: Relative extinction coefficients as a function of wavelength in four regions with Galactic longitude ranging from $0\le l \le360$, and range of Galactic latitude indicated in the legend.}
\label{cr_los}
\end{figure}

We used the same Bayesian approach outlined in \S4.1 to measure the relative extinction coefficients for all 10 bands in each unique region. In Figure \ref{cr_los} we show the relative extinction coefficients for the lines of sight indicated by the figure legends.  The error bars shown indicate the standard deviation about the median $A_\lambda$ slopes from each 5,000 iteration MCMC chain.

Significant variation in the shape of the extinction law for $H$, $K_s$ and WISE bandpasses was seen between different regions in both Galactic latitude and longitude in Figure \ref{cr_los}. In most regions the deviation of the relative extinction law in the IR from that of the median values from \S4.1 are all in the same direction. We emphasize that the dramatic changes in the extinction between each LoS primarily manifested themselves in the IR, while the optical bands remained consistent with $R_V\sim3.0-3.1$ from \citet{berry2012}.

\begin{figure}
\centering
\includegraphics[width=3.5in]{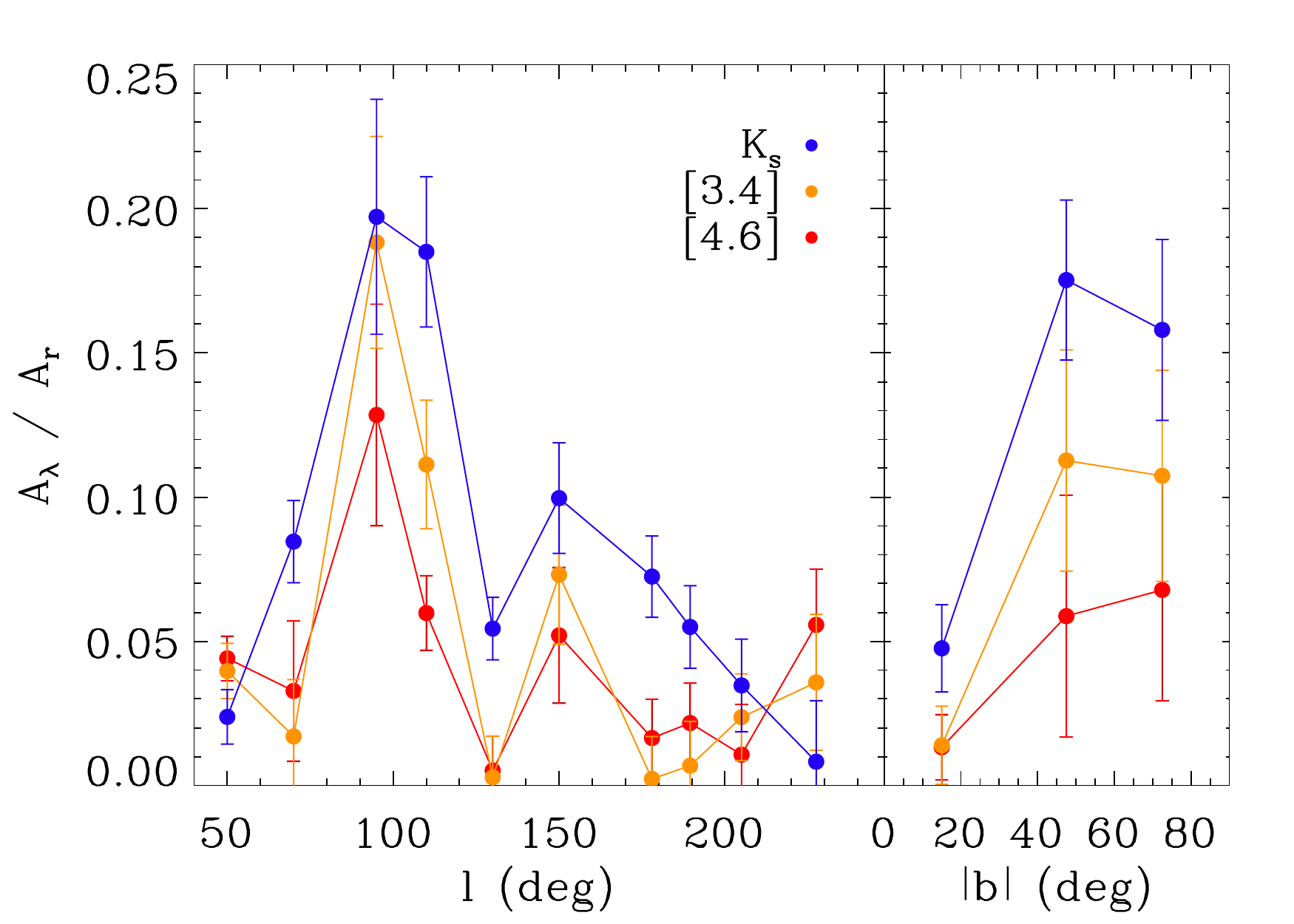}
\caption{Variations with Galactic longitude (left) and latitude (right) in the fractional extinction coefficients for the $K_s$, W1, and W2-bands.}
\label{lb}
\end{figure}

There is also correlation between Galactic position and the steepness of the extinction law. Figure \ref{lb} shows the fractional extinction coefficients versus Galactic coordinates for three infrared bands. The trend with Galactic longitude was weaker. \citet{gao2009} measured extinction variations with Galactic longitude in the infrared, indicating a correlation with spiral arm structure. Their overall trend of decreasing infrared extinction towards the Galactic anti-center matches our observation here. 
A strong coherent trend of increasing fractional infrared extinction with Galactic latitude was observed. Using a small sample of stars with a limited range in optical extinction, \citet{larson2005} demonstrated infrared extinction variations relative to the $V$-band as a function of Galactic latitude. They found a steeper extinction curve with increasing latitude, implying the existence smaller dust grain sizes with height above the plane. Our results seem to strongly contradict this finding, and may instead imply a difference in grain chemistry with Galactic latitude.

\section{Color Outliers}
\label{sec:dust}

Using our initial sample of low extinction point sources, described in \S\ref{sec:data}, we carried out a  search for objects with unusually red colors in the WISE passbands. For this preliminary investigation we limited our search to the WISE preliminary data release footprint, which reduced the number of sources to check by eye. Many studies have utilized combinations of optical and mid IR passbands to find ``excess'' IR emission, which is frequently attributed to a surrounding dust shell or disk for stellar objects. For example, \citet{beichman2006} determined the mean $K-[24]$ color as a function of spectral type for nearby FGKM stars in order to search for excess emission from disks. The mid-IR dust emission is apparent in this $K-[24]$ color, with $\Delta(K-[24])$ up to 1 magnitude larger than the intrinsic stellar values \citep{gorlova2006}. Similarly, we have used our locus to find stars with excess emission in the the $K_s-W3$ color space.

We first selected all sources from our low-extinction sample that had robust $W3$-band measurements, requiring a photometric error of $\sigma_{W3}\le 0.2$. We also required stars to have $W2<12$ and $W3<11$ to remove objects with low signal-to-noise in the WISE bands. This resulted in a refined subsample of 4892 objects to search for outliers in the WISE bands. 
 
\begin{figure}
\centering
\includegraphics[width=3.5in]{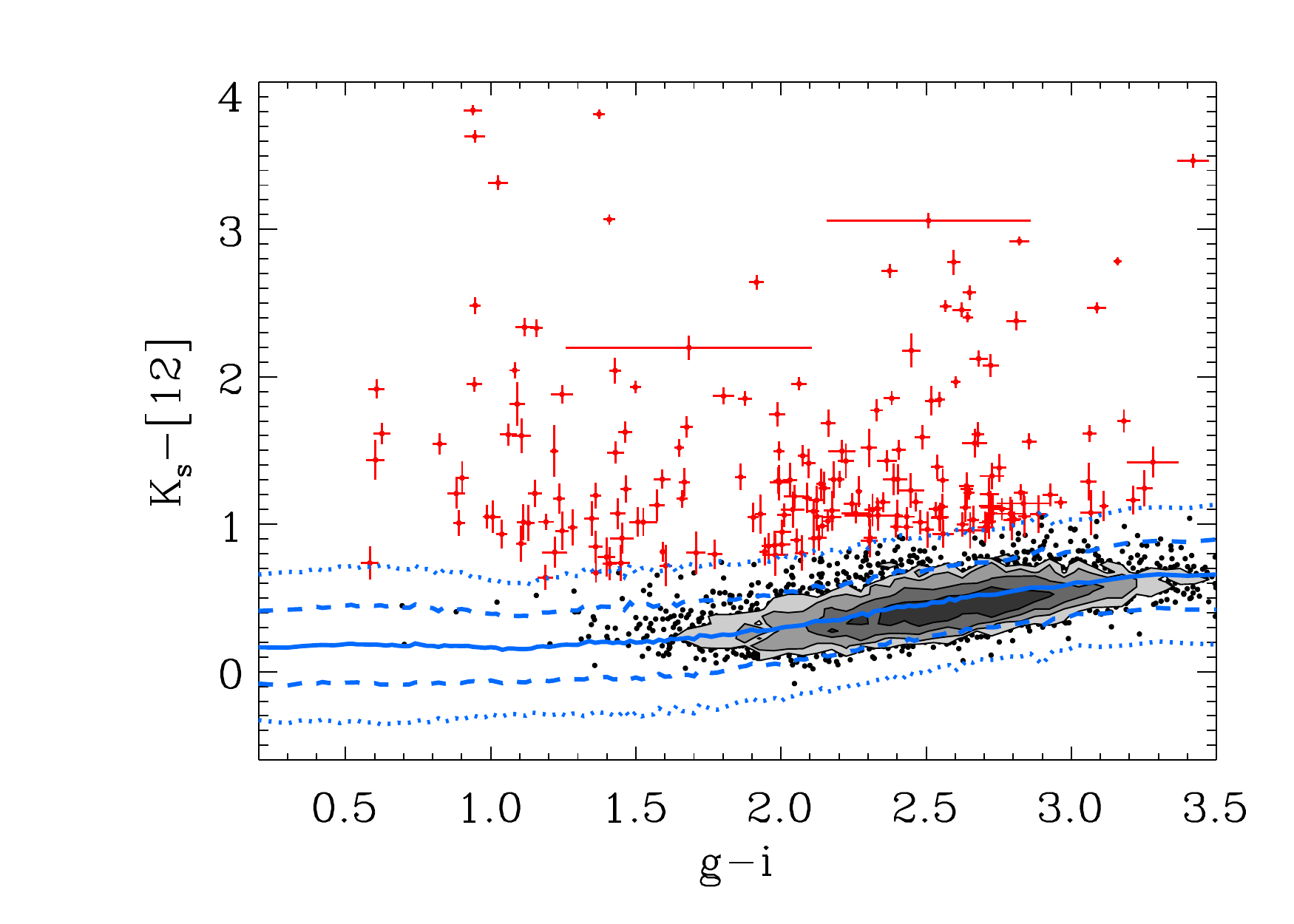}
\caption{Optical versus infrared color-color diagram of the refined subsample of 4,892 objects, as defined in the text. Objects selected for their $K_s-W3$ excess (red crosses) are shown with their photometric error bars. The stellar locus in this color space (solid blue line) traces the high density contours of stars. The 1- and 2-$\sigma$  uncertainties on the locus are also shown (dashed and dotted blue lines, respectively).}
\label{dust1}
\end{figure}

From this subsample, we selected objects with $K_s-W3$ colors more than two standard deviations redder than the locus, as demonstrated in Figure \ref{dust1}. We also required objects to have $K_s-W3<4$, to avoid contamination from objects whose WISE-only colors show them to be quasars \citep{wise,wu2012}. There were 126 such objects that were not removed by our earlier $u-g$ cut, but were eliminated by our $K_s-W3<4$ requirement. 
The final sample of color outliers contained 199 objects, highlighted in red in Figure \ref{dust1}.

We elected to not search using the W4-band, as most objects in the 4892 subsample had poor signal-to-noise in that filter. The significant number of quasars at red $K_s-W3$ colors that were removed highlights the difficulty in separating genuine dust disks from contaminating AGN using only optical and NIR filters, and the great utility of matching sources to WISE for classification.

\citet{kennedy2012} found about 4\% of the $\sim$180,000 objects surveyed in the Kepler field \citep{borucki2010} had apparent W3- or W4-band excess emission in WISE. Most of these were attributed to spurious emission from dust over-densities, with only $\sim$0.15\%  deemed real mid-IR excesses. Such a low rate may also be due to chance alignments of background galaxies, rather than bona fide dust emission from stars in the Galactic disk.

Our gross rate of detecting excess emission candidates was 199 out of 4892 objects, or $4.1\pm0.3$\%, quite close to the initially detected rate of 4\% found by \citet{kennedy2012} in the Kepler field. If we assume the same efficiency in finding bone fide dust disks as in \citet{kennedy2012}, we would expect only $\sim$7 systems to host astrophysically real infrared excess. 
 
\begin{figure}
\centering
\includegraphics[width=3.5in]{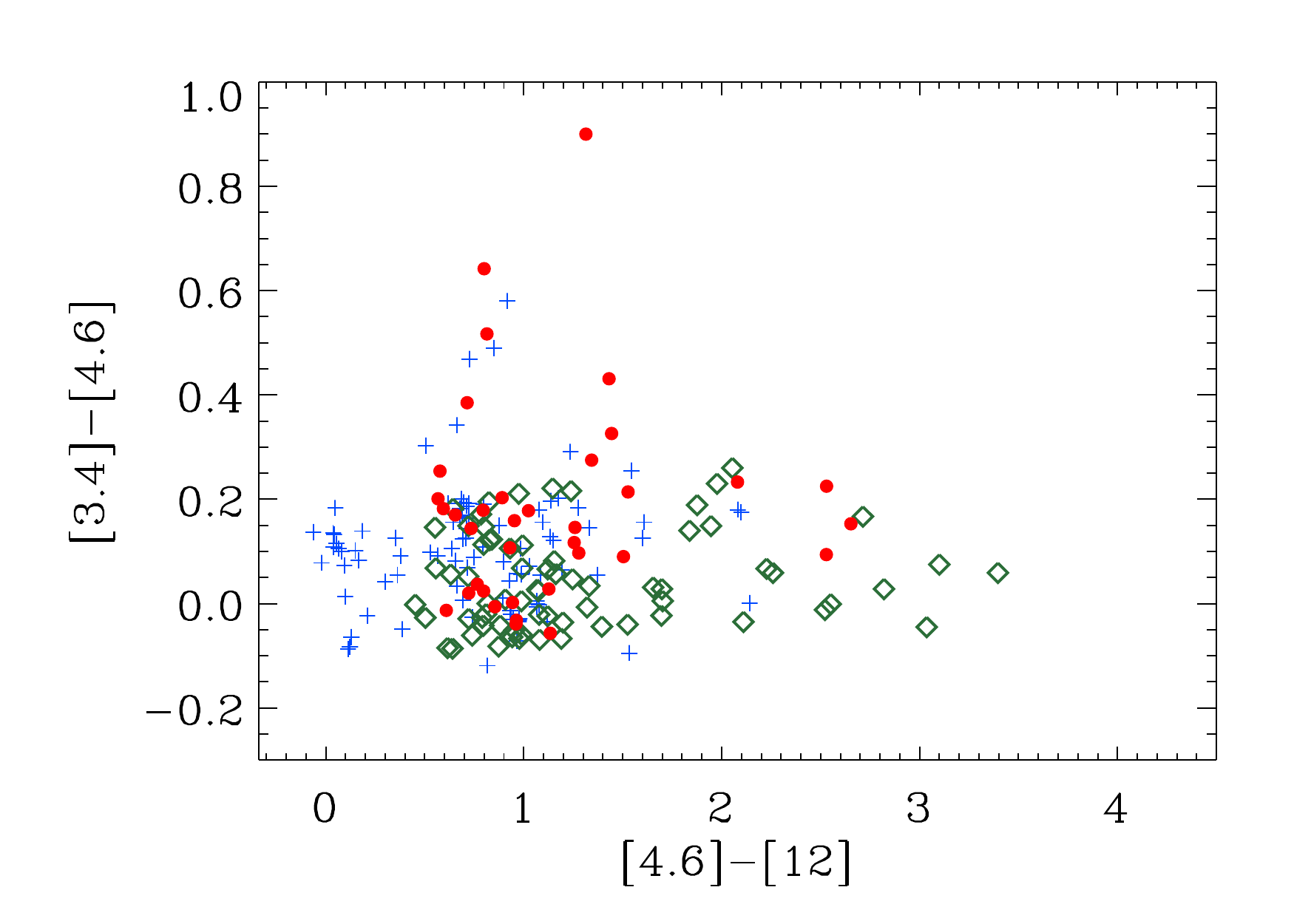}
\caption{WISE only color-color space for the 199 infrared excess candidates. SDSS imaging for each source has been manually inspected. Candidates were put in three categories: objects with a neighboring star within 10'' (blue crosses), objects with contaminating background galaxies (green diamonds), and objects that appeared isolated within 10'' (red circles).}
\label{dust4}
\end{figure}

The primary sources of contamination were nearby stars that may not have been properly separated in the WISE or 2MASS imaging, or background galaxies. To further investigate the contamination for our 199 infrared excess candidates, we manually inspected the SDSS imaging within a radius of 20'' around each target. We manually classified the infrared excess candidates in to three categories: 70 objects with background galaxies whose extent was within 10'' of the target, 93 objects with resolved neighboring point sources within 10'' of the target, and 36 objects with no discernible contaminant with 10''. The WISE  $(W2-W3, W1-W2)$ color-color distribution for each category is shown in Figure \ref{dust4}. 

About half of infrared excess candidates whose SDSS imaging showed them to have nearby stars fell close to the stellar locus in the WISE-only color-color diagram, blue crosses centered at approximately $(0.2,0.1)$ in Figure \ref{dust4}. Objects with clear contamination due to background galaxies in Figure \ref{dust4} inhabit a large range in $W2-W3$ color, red ward of the stellar locus, shown as green diamonds. These background galaxy contaminants included both small galaxies that, as with the neighboring star, were not resolved properly in the WISE imaging, and galaxies with highly extended structure that overlapped with the foreground target star and disrupted the local background determination in WISE photometry. Some of the systems with contaminating nearby stars (blue crosses) may have actually been contaminated by unresolved background galaxies instead, creating overlap between these two types of systems in the WISE color-color diagram.

A small number of objects follow a linear trend away from the stellar locus, which lie along the asymptotic giant branch track in the WISE color-color space ($W2-W3\sim1$). Of the objects with no discernible contamination (red circles), only six had $W1< 11$ mag. Faint objects with $W1>11$ have a much higher likelihood of being background galaxy contaminants due to increasing densities of galaxies and AGN at fainter WISE magnitudes \citep[][Nikutta 2014 in prep]{yan2013}. These six objects are the most likely warm dust disk candidates in our sample, placing a tentative fraction of stellar systems with real infrared excess at $0.12\pm0.05\%$, where the uncertainty quoted is the binomial error.  These may also be due to unresolved contamination or binary systems. These three categories of objects (galaxy contaminants, stellar contaminants, and dust disk candidates) all spanned the full range of $(g-i)$ color in Figure \ref{dust1}. The six most likely dust candidates, however, all had $(g-i) > 2.5$, indicating the apparent over-density of dust candidates in Figure \ref{dust1} at blue optical colors was solely due to contamination. Confirming the presence of dust and characterizing the dust temperatures for these candidates will require followup at wavelengths of 50--100 $\mu$m.

\section{Summary}
\label{sec:conclusion}

We have presented a study of the fundamental properties of stars across a wide range in wavelength. Our stellar locus, derived from a million low-extinction sources, will be of great utility to many future studies with the powerful multi-wavelength combination of SDSS - 2MASS - WISE. Spectroscopically confirmed low-mass stars and brown dwarfs from SDSS, matched to WISE photometry, will also provide an extension of our color locus to lower mass objects (S. J. Schmidt 2014 in prep).

A brief summary of our work is as follows: 

\begin{enumerate}
\item A measurement of the 10-dimensional color locus was presented, from SDSS $u$-band to WISE $W3$-band, using 1,052,793 stars with low extinction $(A_r<0.125)$. This locus contains the best characterization of stellar colors in WISE passbands to date.
\item We have empirically measured the $r$-band relative dust extinction coefficients, $A_\lambda/A_r$, for each of the photometric bands in our sample, providing strong constraints for dust composition models in the infrared.
\item Variations in the infrared dust extinction have been shown for different lines of sight. Coherent trends with both Galactic latitude and longitude were seen. Increasing relative infrared extinction with increasing Galactic latitude was found to be in opposition to previous observations. A detailed follow-up investigation of the properties of dust extinction and emission in the infrared with WISE is strongly motivated. 
\item From a subset of our sample we recovered 199 infrared excess candidates that span a wide range of optical colors. The majority of these were found to be contaminants from neighboring stars or background galaxies. Six objects appear to be bona fide infrared excess systems, possibly indicative of dust disks. Higher resolution and longer wavelength followup is required to verify these systems.
\end{enumerate}

\section*{Acknowledgements}

We graciously thank the anonymous referee for whose comments significantly clarified and improved this manuscript. 
JRAD and ACB acknowledge support from NASA ADP grant NNX09AC77G. \v{Z.} Ivezi\'{c} acknowledges support by NSF grants AST-0707901 and 
AST-1008784 to the University of Washington, and by NSF grant
AST-0551161 to LSST for design and development activity. 

Funding for SDSS-III has been provided by the Alfred P. Sloan
Foundation, the Participating Institutions, the National Science
Foundation, and the U.S. Department of Energy Office of Science.
The SDSS-III web site is http://www.sdss3.org/.
SDSS-III is managed by the Astrophysical Research Consortium for the
Participating Institutions of the SDSS-III Collaboration including the
University of Arizona,
the Brazilian Participation Group,
Brookhaven National Laboratory,
University of Cambridge,
Carnegie Mellon University,
University of Florida,
the French Participation Group,
the German Participation Group,
Harvard University,
the Instituto de Astrofisica de Canarias,
the Michigan State/Notre Dame/JINA Participation Group,
Johns Hopkins University,
Lawrence Berkeley National Laboratory,
Max Planck Institute for Astrophysics,
Max Planck Institute for Extraterrestrial Physics,
New Mexico State University,
New York University,
Ohio State University,
Pennsylvania State University,
University of Portsmouth,
Princeton University,
the Spanish Participation Group,
University of Tokyo,
University of Utah,
Vanderbilt University,
University of Virginia,
University of Washington,
and Yale University.

This research has made use of the NASA/ IPAC Infrared Science Archive, which is operated by the Jet Propulsion Laboratory, California Institute of Technology, under contract with the National Aeronautics and Space Administration.

This publication makes use of data products from the Two Micron All Sky Survey, which is a joint project of the University of Massachusetts and the Infrared Processing and Analysis Center/California Institute of Technology, funded by the National Aeronautics and Space Administration and the National Science Foundation.	

This publication makes use of data products from the Wide-field Infrared Survey Explorer, which is a joint project of the University of California, Los Angeles, and the Jet Propulsion Laboratory/California Institute of Technology, funded by the National Aeronautics and Space Administration.


\begin{table*}
\caption{The optical and NIR color locus as a function of $g-i$ color bins. The number of stars in each bin is included (\#). Values in parenthesis indicate the standard deviation of the locus in each color. The entire table is available online in machine readable format.}
\label{bigtable1}
\begin{tabular}{@{}lcccccccc}
\hline
$g-i$ & \# & $u-g$ & $g-r$ & $r-i$ & $i-z$ & $z-J$ & $J-H$ & $H-K_s$ \\
\hline
 0.200 &     5708 &  1.087 ( 0.032) &  0.268 ( 0.029) &  0.082 ( 0.021) & -0.007 ( 0.018) &  0.772 ( 0.034) &  0.245 ( 0.041) &  0.039 ( 0.045)\\
 0.300 &     4264 &  1.077 ( 0.066) &  0.303 ( 0.025) &  0.098 ( 0.017) &  0.010 ( 0.020) &  0.790 ( 0.033) &  0.269 ( 0.040) &  0.044 ( 0.046)\\
 0.350 &     9259 &  1.082 ( 0.052) &  0.327 ( 0.026) &  0.118 ( 0.022) &  0.018 ( 0.021) &  0.807 ( 0.034) &  0.289 ( 0.042) &  0.044 ( 0.046)\\
 0.400 &    15934 &  1.075 ( 0.042) &  0.374 ( 0.031) &  0.132 ( 0.021) &  0.031 ( 0.019) &  0.829 ( 0.036) &  0.314 ( 0.043) &  0.048 ( 0.046)\\
 0.420 &     2494 &  1.112 ( 0.029) &  0.366 ( 0.029) &  0.133 ( 0.024) &  0.010 ( 0.020) &  0.829 ( 0.034) &  0.300 ( 0.043) &  0.047 ( 0.045)\\
\hline
\end{tabular}
\end{table*}

\begin{table*}
\caption{The mid-IR color locus as a function of $g-i$ color. The number of stars in each bin from the primary sample is included (\#), as well as the $W3$-limited subsample (\#$_{W3}$). Values in parenthesis indicate the standard deviation of the locus in each color. The entire table is available online in machine readable format.}
\label{bigtable2}
\begin{tabular}{@{}lccccc}
\hline
$g-i$ & \# & $K_s-W1$ & $W1-W2$ & \#$_{W3}$ & $W2-W3$\\
\hline
 0.20 &     5708 &  0.032 ( 0.047) & -0.016 ( 0.024) &  0 & \ldots (\ldots) \\
 0.30 &     4264 &  0.033 ( 0.047) & -0.017 ( 0.024) &  0 & \ldots (\ldots) \\
 0.35 &     9259 &  0.035 ( 0.046) & -0.017 ( 0.023) &  0 & \ldots (\ldots) \\
 0.40 &    15934 &  0.038 ( 0.047) & -0.018 ( 0.023) &  0 & \ldots (\ldots) \\
 0.42 &     2494 &  0.035 ( 0.045) & -0.021 ( 0.022) &  0 & \ldots (\ldots) \\
\hline
\end{tabular}
\end{table*}

\begin{table}
 \caption{Median extinction coefficients relative to the $r$-band.}
 \label{extinct_table}
 \begin{tabular}{@{}clll}
  \hline
Filter & $\lambda$ & $A_\lambda/A_r$ & $\sigma(A_\lambda/A_r)$\\
  \hline
$u$ & 0.335 $\mu$m & 1.83 & 0.02 \\
$g$ & 0.469 $\mu$m & 1.39 & 0.01 \\
$r$ & 0.617 $\mu$m & 1 & \ldots \\
$i$ & 0.748 $\mu$m &  0.76 & 0.01 \\
$z$ &  0.893 $\mu$m & 0.53 & 0.02 \\
$J$ & 1.24 $\mu$m & 0.30 & 0.03 \\
$H$ & 1.66  $\mu$m & 0.21 & 0.03\\
$K_s$ & 2.16 $\mu$m & 0.15 & 0.03 \\
$W1$ & 3.35 $\mu$m & 0.09 & 0.03 \\
$W2$ & 4.46 $\mu$m & 0.05 & 0.04 \\
$W3$ & 11.6 $\mu$m & 0.13 & 0.05 \\
 \hline
 \end{tabular}
 \end{table}

\bsp

\label{lastpage}
\end{document}